\documentclass{article}
\usepackage{braket}

\usepackage[english]{babel}

\usepackage[letterpaper,top=2cm,bottom=2cm,left=3cm,right=3cm,marginparwidth=1.75cm]{geometry}

\usepackage{amsmath}
\usepackage{graphicx}
\usepackage{amssymb}
\usepackage{braket}
\usepackage{float}
\usepackage{svg}
\usepackage{nicefrac}
\usepackage{authblk}
\usepackage{hyperref}
\usepackage{mathtools}
\usepackage{comment}

\title{Optical spin tomography in a telecom C-band quantum dot}
\author[1,$\dagger$,$*$]{Junyang Huang}
\author[1,$\dagger$]{Ginny Shooter}
\author[1,2,$\dagger$]{Petros Laccotripes}
\author[1]{Andrea Barbiero}
\author[2]{David A. Ritchie}
\author[1]{Andrew J. Shields}
\author[1]{Tina M\"{u}ller}
\author[1]{R. Mark Stevenson}
\affil[1]{Toshiba Europe Limited, 208 Cambridge Science Park, Milton Road, Cambridge, CB4 0GZ, UK}
\affil[2]{Cavendish Laboratory, University of Cambridge, JJ Thomson Avenue, Cambridge, CB3 0HE, UK}
\affil[$*$]{\textit{Email:} junyang.huang@toshiba.eu}
\affil[$\dagger$]{These authors contributed equally.}
\date{}

\usepackage{setspace}
\begin{document}
\maketitle

\begin{abstract}
\noindent
A central challenge for scalable quantum networks is the realization of coherent interfaces between stationary qubits and telecom-band photonic qubits for long-distance entanglement distribution. Semiconductor quantum dots emitting at telecom wavelengths present a promising spin-photon platform, and a precise understanding of the properties of the confined spin is crucial for optimising its interplay with the photonic qubit. Here, we simultaneously benchmark the electron and hole $g$-factors and coherence properties of a droplet epitaxy QD, solely from time and polarization resolved photon correlations. These measurements identify the hole as the preferable qubit for spin-photon entanglement in quantum network nodes. We then perform full state tomography of the confined hole ground state to reveal subtle anisotropies in the spin precession, providing essential diagnostics for minimizing phase errors critical for deterministic multiphoton entanglement generation.

\end{abstract}

% \tableofcontents

\section*{Introduction}

Future quantum networks are envisioned to process quantum information across many remote quantum registers connected via shared entanglement \cite{azuma2023quantum}, with transformative applications in distributed quantum computing \cite{main2025distributed} and non-local quantum sensing and metrology \cite{kim2024distributed, zhao2021field}. To achieve long-distance entanglement distribution, all-photonic approaches to quantum repeaters have been put forward \cite{azuma2015all, li2019experimental}. All-photonic repeaters avoid the need for long-lived quantum memories and thus do not require global synchronization of entanglement generation across the full network, relying instead on local operations on pre-prepared cluster states. Linear-optical fusion can assemble large cluster states from probabilistic sources, yet its inherent nondeterminism makes the scheme overhead-heavy. A leading alternative is deterministic sequenctial generation from a coherently controlled stationalry qubit, as in the Lindner–Rudolph protocol\cite{lindner2009proposal}. In such emitter-based implementations, the crucial primitive is a coherent spin-photon interface, ideally operating at telecom wavelengths, where loss in standard optical fibers is minimal.
%with transformative applications in quantum key distribution \cite{xu2020secure, pittaluga2025long}, distributed quantum computing\cite{main2025distributed, maring2024versatile}, 
%and non-local quantum sensing and metrology \cite{kim2024distributed, zhao2021field}.
%Realising this vision of scalable quantum networks requires robust quantum nodes that interface stationary qubits with flying photonic qubits over long distances \cite{kimble2008quantum}, making it critical to establish faithful coherent spin-photon interfaces at telecom wavelengths, where optical fiber losses are minimal. 

Semiconductor quantum dots (QDs) can confine single electron or hole spins, whose well-defined optical selection rules enables precise initialization, coherent control, and spin state readout. Combined with coherence times sufficient for multiphoton protocols and a strong magneto-optical response, QD spin qubits have been established as a desirable solid-state platform for deteriministic generation of multiphoton cluster states \cite{lindner2009proposal, schwartz2016deterministic, appel2022entangling, cogan2023deterministic, coste2023high}. %\cite{burkard2023semiconductor,jackson2021quantum, ding2019coherent,benny2011coherent}. 
A detailed understanding of coherent spin dynamics, including spin precession and phase evolution, is crucial for realizing high-fidelity spin-photon entanglement \cite{togan2010quantum, coste2023high, cogan2023deterministic, laccotripes2024spin} and robust spin-based quantum memory operations \cite{appel2025many}.

To enable compatibility with existing fiber-optic infrastructure, it is essential to realize these capabilities at telecom wavelengths (1310 nm and 1550 nm). Recent advances in epitaxial growth and material engineering have extended the emission wavelength of III–V QDs into this regime, demonstrating single-photon emission and coherent spin control directly at C-band wavelengths \cite{dusanowski2022optical, laccotripes2024spin, Peniakov.29042025}, while integration into nanophotonic structures has provided Purcell enhancement and polarization control \cite{Nawrath.2023, Holewa.2024, barbiero2024polarization, barbiero2025purcell, hauser2025deterministic}. Yet, achieving spin–photon interfaces at telecom with the same performance as at sub-$\mu$m wavelengths remains a major challenge due to the increased charge noise and the less mature nanophotonic device engineering. Although some steps toward spin-state characterization in this regime have been reported \cite{Peniakov.29042025}, full quantum-state tomography of a spin in a telecom C-band QD has not been achieved.

% Recent progress in epitaxial growth and material engineering has successfully extended III-V QDs into the telecom spectral regime, showcasing single-photon emission and coherent spin control directly at O-band and C-band wavelengths \cite{laccotripes2024spin, wells2023coherent}, while integration into telecom nanophotonic structures has led to Purcell enhancement and improved polarization control \cite{barbiero2024polarization, barbiero2025purcell, hauser2025deterministic}. A detailed understanding of coherent spin dynamics, including spin precession and phase evolution, is crucial for realizing high-fidelity spin-photon entanglement \cite{coste2023high, cogan2023deterministic, laccotripes2024spin} and robust spin-based quantum memory operations \cite{appel2025many}. Nevertheless, full quantum state characterization, especially spin-state tomography in this wavelength range, remains underdeveloped. Current tomography methods typically rely on electrical or microwave control, complicating their scalability and integration into photonic platforms. Moreover, resonant excitation schemes that depend on polarization filtering are difficult to implemented in common nanophotonic structures such as circular Bragg gratings \cite{rickert2024high}. Therefore, demonstrating a fully optical, and polarization filtering free approach to spin control and tomography in telecom-band QD represents a crucial step in the development of practical, fiber-compatible quantum network nodes.

Crucially, deterministic multiphoton entanglement in the Lindner–Rudolph scheme \cite{lindner2009proposal}, the target application, imposes clear design criteria. The resident ground-state spin should maximize its inhomogeneous coherence time \(T_{2}^{*}\), so that phase errors accumulate slowly over many optical cycles, while the excited-state spin should precess much more slowly than it radiatvely decays, \(T_{\text{prec}}^{\text{exc}} \gg \tau_{\text{rad}}\). A pronounced separation between excited- and ground-state precession rates further suppressses spin-flip decay. Assessing 
$g$-factors and coherence times for both spin species is therefore a prerequisite for choosing the optimal entangler.

Here, we first develop a method to simultaneously evaluate basic spin characteristics, i.e. the coherence times and g-factors, of both the ground- and excited-state spin in a single polarisation- and time-resolved correlation measurement. Building on the approach introduced in \cite{PhysRevB.101.035424}, we employ a fully optical and polarization-filtering-free approach to reconstruct the complete quantum state of a heavy-hole spin in a telecom C-band InAs/InP droplet epitaxy QD. Using phonon-assisted (PA) excitation pulses in complementary linear polarizations, we map the instantaneous ground-state Bloch vector onto the polarization statistics of the sequentially emitted photons. Time-tagged correlation histograms then reveal all spin projections without applying microwaves, electrical manipulation, or resonant laser filtering.

\begin{figure}[H]
\centering
\includegraphics[width=1\textwidth]{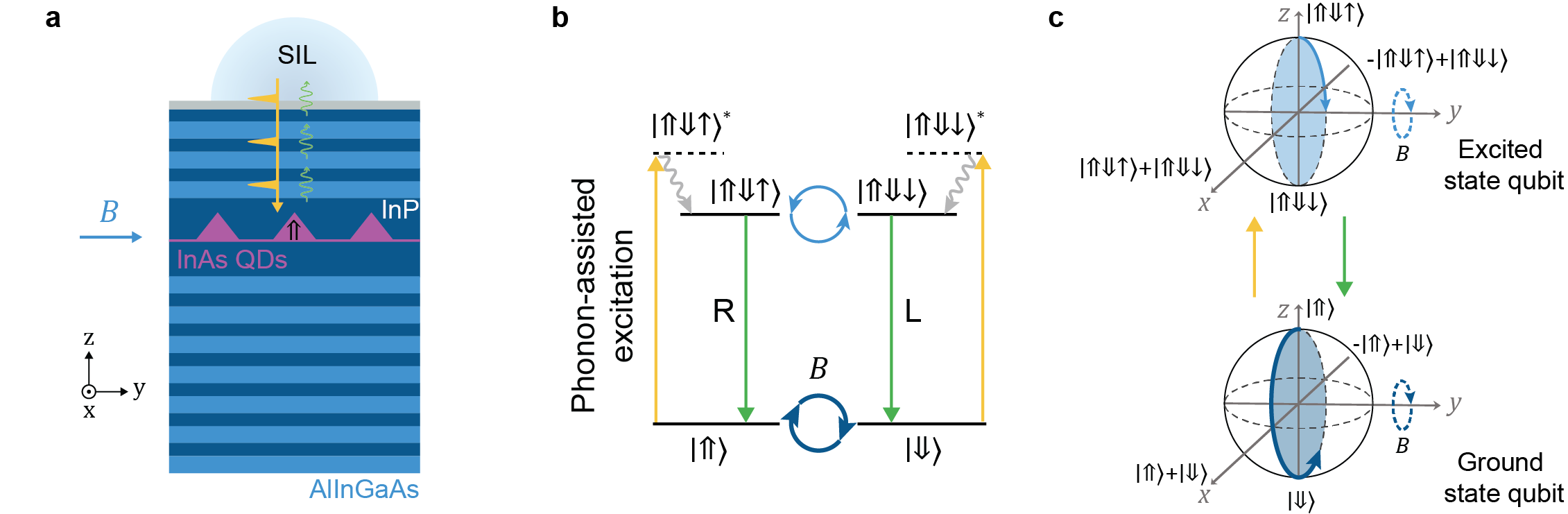} 
\caption{\textbf{Device architecture and hole-trion transition under phonon-assisted excitation.} (a) Schematics of the QD-microcavity device. Droplet epitaxy InAs QDs (purple) are embedded within a distributed Bragg reflector structure composed of InP/AlInGaAs, and capped with a solid immersion lens (SIL) to enhance both pulsed laser excitation and photon collection efficiency. The QD growth axis defines the \(z\) direction, with a weak magnetic field applied along the \(y\) axis (Voigt geometry). (b) Polarization
selection rules governing the hole-trion optical transitions under coherent phonon-assisted excitation. (c) Bloch sphere depiction of spin qubits in the ground and excited states. The spin-up and spin-down eigenstates are aligned along the \(z\) axis, while coherent superposition states reside in the \(x-y\) plane. Blue arrows illustrate the spin precession of the trion and hole states within the \(x-z\) plane under the magnetic field.}
\label{fig:intro} 
\end{figure} 

The droplet epitaxy InAs/InP QDs investigated here are grown within an InP/AlInGaAs distributed Bragg reflector cavity structure (\autoref{fig:intro}a). A solid immersion lens is mounted atop the sample to enhance both the excitation efficiency and the photon collection rate. The QD used in this study supports a positively charged exciton ($\text{X}^{+}$) emitting in the telecom C band ($\sim$1534 nm), realizing an optical $\pi$-system in which the heavy-hole ground states $\left|\Uparrow\right\rangle$ and $\left|\Downarrow\right\rangle$ are coherently linked to the excited-state electron in trion configurations $\left|\Uparrow\Downarrow\uparrow\right\rangle$ and $\left|\Uparrow\Downarrow\downarrow\right\rangle$ (\autoref{fig:intro}b). Spin-preserving excitation is achieved with a PA scheme, where a $\sim$15 ps pulse is blue-detuned above the optical resonance and the trion state is populated via coupling with a longitudinal-acoustic phonon \cite{glassl2013proposed, ardelt2014dissipative, reiter2012phonon}. This approach eliminates laser leakage at the detection wavelength via spectrally filtering, and, crucially, preserves the phase coherence between the ground state spin components during excitation. The transitions follow circular polarization selection rules: recombination of the trion $\left|\Uparrow\Downarrow\uparrow\right\rangle$ ($\left|\Uparrow\Downarrow\downarrow\right\rangle$) generates an R- (L-) polarized photon, directly projecting the emitter back into the $\left|\Uparrow\right\rangle$ ($\left|\Downarrow\right\rangle$) ground state.

We apply a weak transverse magnetic field along the \(y\)-direction (Voigt geometry) to lift the degeneracy of both the ground- and excited-state spin manifolds. The resultant Larmor precessions of the ground state hole and the photoexcited electron are depicted in \autoref{fig:intro}c in the Bloch sphere representation, where the spin eigenstates align along the \(z\)-axis and the coherent superpositions span the equatorial \(x-y\) plane. During the ground-state residence time, the hole spin precesses within the \(x-z\) plane. Following optical excitation, the system transitions to the trion state, where the excited electron spin can inherit the phase of the hole, and continues to precess prior to radiative recombination. Crucially, such preservation of coherence during the transition enables phase-sensitive readout of the spin's trajectory and underpins the tomography protocol in this work.

\section*{Ground and excited state spin precession and coherence}

\begin{figure}[H]
\centering
\includegraphics[width=1\textwidth]{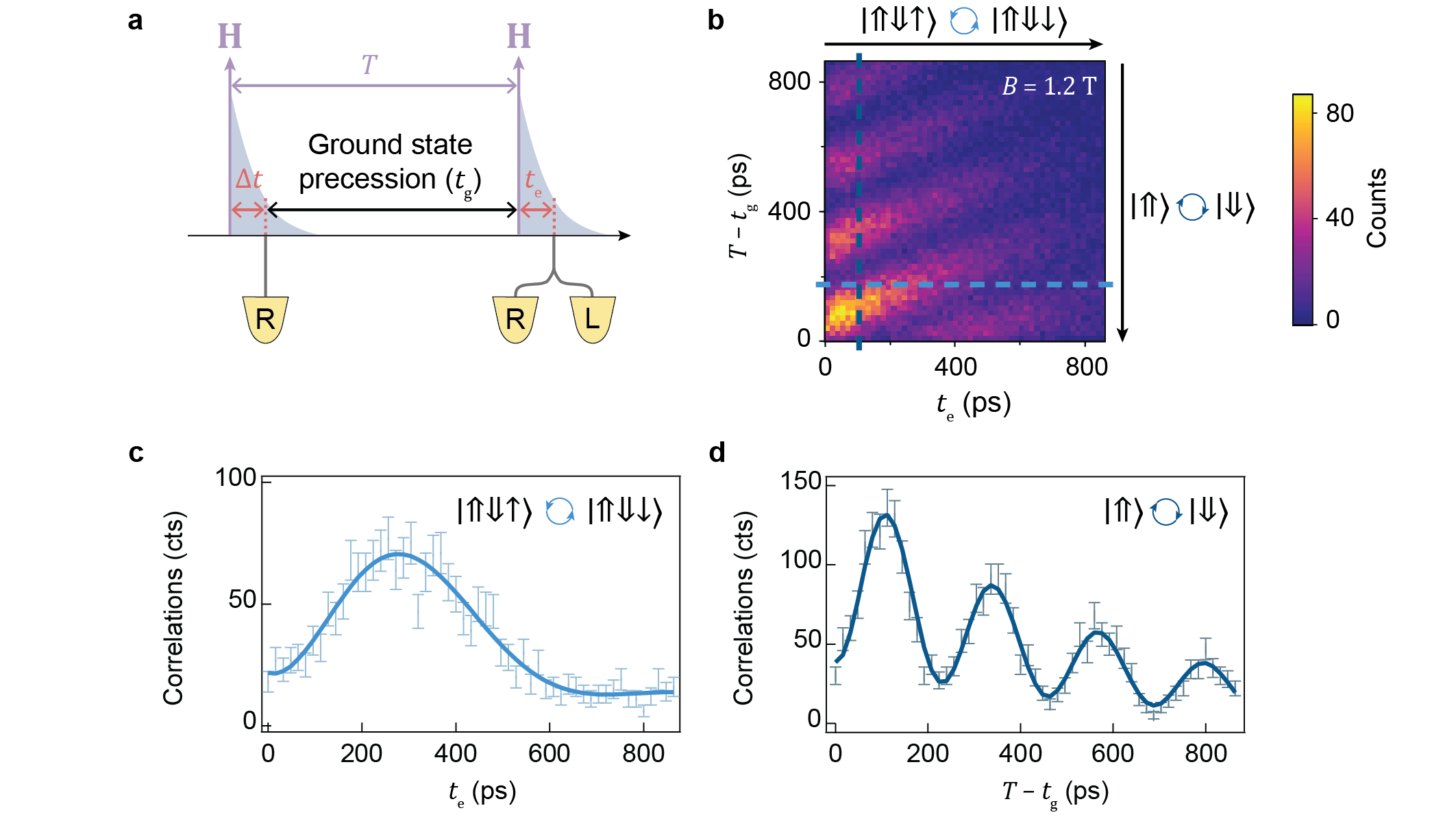} 
\caption{\textbf{Spin precession dynamics in ground and excited states.} (a) Experimental scheme illustrating the polarized optical excitation pulse sequence (purple arrows) and detection configuration to measure emitted photons in the circular basis, using electronic polarization controllers and polarizing beamsplitters. (b) Time-resolved, two-photon correlation map for revealing spin dynamics for both ground and excited states at $B$ = 1.2 T. The two-dimensional histogram displays correlated photon arrival times, measured in orthogonal circular polarization channels (R-L). Horizontal axis shows $t_{e}$, the time spent in the excited state after the second excitation pulse. $\Delta$$t$ on the y-axis denotes the time spent in the excited state after the first excitation pulse, complementary to the ground-state precession time $t_{g}$ (i.e. $T$ = $\Delta$$t$ + $t_{g}$). (c) Temporal cross-section along a 32-ps-high horizontal slice (dashed line in (b)), isolating the slow excited-state electron spin precession, with theoretical fit shown as a solid line. (d) Corresponding 32-ps-width vertical slice of (b), revealing the fast ground-state hole spin precession.The binning size in (c) and (d) is 16 ps.}
\label{fig:ground_and_exc} 
\end{figure}

To realise spin tomography, we first assess the spin dynamics of our telecom-band QD, using a two-photon correlation measurement. \autoref{fig:ground_and_exc}a depicts the laser-pulse sequence and detection scheme and here we focus on a hole-trion transition as an example. Two H-polarised, PA pulses, separated by time $T$, are applied in each repetition cycle. Detection of an R-polarised photon at $\Delta$$t$ after the first pulse heralds preparation of the ground-state spin in $\ket{\Uparrow}$. During the subsequent delay ($t_{g}$) until the arrival of the next excitation pulse, the hole Larmor-precesses in the \(x-z\) plane under the applied magnetic field. By post-selecting events along the exponential tail with different $\Delta$$t$, we map different precession times of the hole spin, before the second excitation pulse. Upon arrival of the second H pulse, the instantaneous ground-state spin—whether in $\ket{\Uparrow}$, $\ket{\Downarrow}$, or a superposition thereof, depending on the elapsed time—is coherently transferred into the trion, thereby imprinting the accumulated phase onto the excited-state electron. The electron then continues to precess for time $t_{e}$ until spontaneous emission returns the system to the ground state. Finally, the second photon is analysed in the R/L basis, completing the spin readout. We operate with an in-plane magnetic field of $B$ = 1.2 T to ensure $>$2$\pi$ spin precession of the excited-state electron is resolved within the $\sim$0.8 ns lifetime window. It is noted that in this case, the applied field does not compromise the selection rules. While large transverse field can open diagonal spin-flip transitions, we measure $>90\%$ degree of circular polarisation (DCP) at both $B$ = 0 and 1.2 T, confirming that the optical optical $\pi$ system is preserved. 

The time-tagged two-photon correlation map, displayed in \autoref{fig:ground_and_exc}b, resolves the Larmor precession of both the ground-state hole and the excited-state electron, encoding their precession periods and lower bounds of the coherence times simultaneously. Electron spin precession manifests as slow oscillations parallel to the x axis in \autoref{fig:ground_and_exc}b, mapped against the excited-state dwell time $t_{e}$. The integrated signal from a representative 32 ps-high horizontal slice (\autoref{fig:ground_and_exc}c) is fitted with 
\begin{equation}
I(t)=\frac{I_{0}}{2}\,e^{-t/T_{\text{1}}}
\Bigl[
  1 + V_{0}\,e^{-(t/T_{2}^{*})^{2}}
      \cos\!\bigl(-\tfrac{2\pi t}{T_{\text{prec}}}+\varphi_{0}\bigr)
\Bigr],
\end{equation}
where $T_{\text{1}}$ is the decay time, $T_{2}^{*}$ the dephasing time, $T_{\text{prec}}$ the precession period, $I_{0}$, $V_{0}$, and $\varphi^{0}$ are initial intensity, visibility, and oscillation phase respectively \cite{PhysRevB.101.035424}. The electron spin g-factor of $g_{e}$=0.10±0.01 and the coherence time of $T_{2}^{*}$=0.7±0.2 ns are extracted from the period and visibility of the oscillations.

The precession of the ground state hole spin can be seen in the fast oscillations along vertical slices of \autoref{fig:ground_and_exc}b (dark dashed line). Longer delays $\Delta$$t$ give the QD less time in the ground state before the second laser excitation pulse arrives. \autoref{fig:ground_and_exc}d shows coincidences within a 32-ps wide vertical slice which are summed across rows. Similarly, a hole g-factor of $g_{h}$=0.24±0.01 can be extracted from the oscillations. This value of the ground state hole g-factor agrees with our independent measurement of the excited state hole spin for the same QD \cite{laccotripes2025entangled}. The ground state hole spin coherence time, which is extracted from the oscillation visibility, is limited by the trion lifetime, thus 7±2 ns is a lower bound. By resolving the precession for both ground- and excited-state spins in a single optical dataset, this protocol benchmarks the coherence and precession, providing a quantitative basis for selecting the qubit species and spin configuration for the intended spin-photon interface.

 % and the ratio $g_{h}/g_{e}\approx$ 2.4 is a useful parameter in evaluating the appropriate entangler for the multiphoton protocol.

\section*{Complete Spin Tomography}

\begin{figure}[H]
\centering
\includegraphics[width=0.5\textwidth]{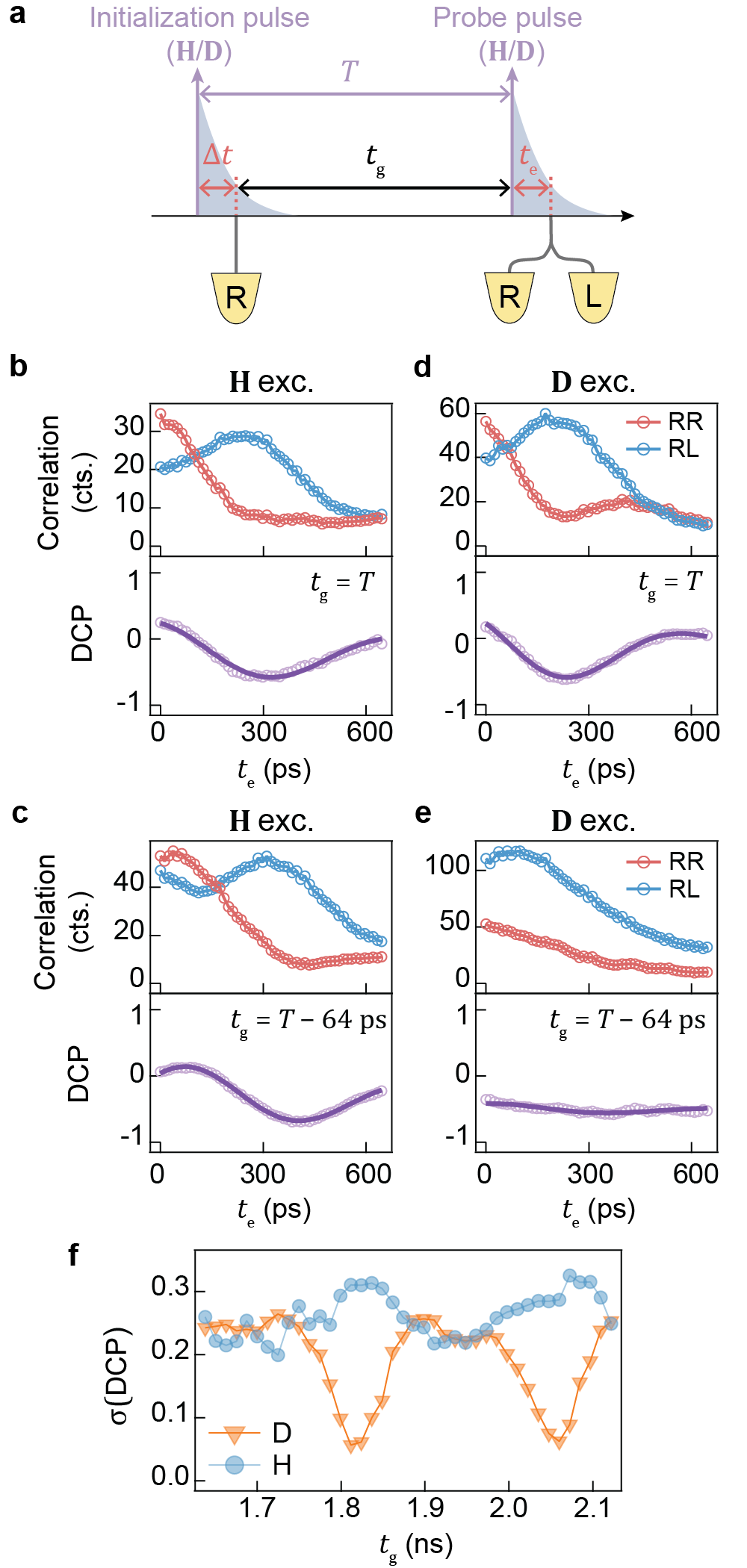} 
\caption{\textbf{Ground state spin evolution and DCP analysis.} (a) Schematic of the excitation pulse sequence and detection setup. The QD is driven by linearly polarized excitation pulses in either H or D polarization, while detection is performed in the circular basis. Ground state precession time is controlled by selecting different temporal slices along the decay of the first excitation (red dashed lines). (b-e) Time-resolved photon correlation measurements for varying ground state precession intervals under H-polarized (b, c) and D-polarized (d, e) excitation, with in-plane magnetic field $B$=1.2 T. In all cases, detection of the first photon is conditioned on R polarization, and the second photon’s detection in R or L polarization allows measurement of the spin’s oscillatory dynamics in the excited state. Panels (b) and (d) correspond to correlations taken at $\Delta$$t$ = $T$\textminus $t_{g}$ = 0 ps, while (c) and (e) represent measurements at 64 ps delay. Red and blue data points denote R-R and R-L coincidence counts, respectively. Lower panels show the extracted DCP from the polarization-dependent correlations, fitted with a damped oscillator model. (f) Standard deviation $\sigma$ of DCP as a function of the ground state precession time under H- and D-polarized excitation.}
\label{fig:HandD} 
\end{figure}

To carry out complete tomography of the ground-state hole spin, we employ the two-pulse protocol illustrated in \autoref{fig:HandD}a, executing it first with two H-polarized and then with two D-polarized excitation pulses. After the heralding first photon with R detection, which projects the QD into the eigenstate $\ket{\Uparrow}$, the hole precesses for an interval $t_{g}$ before the second excitation. The choice of linear polarisation of the probe pulse provides two complementary projections between the evolving ground-state Bloch vector \(\boldsymbol{S} = (S_x, S_y, S_z)\) and the excited state vector \(\boldsymbol{S^{*}}\) which governs the polarization outcome of the second photon, here used as the probe photon:
\begin{equation}
\boldsymbol{S^{*}} =
\left\{
\begin{aligned}
& (S_x,\, S_y,\, S_z),\;\text{H excitation}, \\
& (-S_y,\, S_x,\, S_z),\; \text{D excitation}.
\end{aligned}
\right.
\end{equation}
A D-polarized pulse carries a $\pi/2$ phase between its left- and right-circular components, rotating the Bloch vector by $90^{\circ}$ about the growth axis. The probe photon is analyzed in both R- and L- polarization bases. This projects the excited-state Bloch vector onto the $z$-axis and converts its Larmor precession into the oscillatory R-R and R-L coincidence signals shown in the upper panels of \autoref{fig:HandD}b-e. To extract the phase information from these traces we calculate the DCP
\[
\text{DCP}(t) = \frac{R(t) - L(t)}{R(t) + L(t)}   ,
\]
where $R(t)$ and $L(t)$ denote the time-resolved coincidence counts in the R and L channels, respectively. The DCP dynamics encapsulate the accumulated phase of the ground-state hole spin, given by
\begin{equation}
\text{DCP}(t)=
A_0\, e^{-(t/T_{2}^{*})^{2}}\;
\cos\!\left(-\frac{2\pi t}{T_{\text{prec}}}+\phi_{0}\right),
\end{equation}
where $A_0$ and $\phi_{0}$ are the initial amplitude and oscillation phase, respectively \cite{PhysRevB.101.035424,cogan2018depolarization}. 

\autoref{fig:HandD}b-e show representative data sets acquired for two ground-state precession times by selecting $t_{g}$ = $T$ and $t_{g}$ = $T$\textminus64 ps. From R-L and R-R coincidences, we extract DCP traces plotted together with fits to the damped-oscillation model. With H-polarized excitation (\autoref{fig:HandD}b,c), the DCP retains a constant visibility for different ground state precession times, reflecting the fact that H excitation yields identical mapping of spin components. In contrast, switching to D polarized excitation (\autoref{fig:HandD}d,e) introduces a $\pi$/2 phase shift of the equatorial Bloch vector, causing the DCP visibility to now depend on the accumulated hole spin phase and therefore oscillate as a function of $t_{g}$. This visibility is quantified using the standard deviation of the DCP trace, $\sigma$(DCP), which is shown versus the ground state precession time $t_{g}$ in \autoref{fig:HandD}f. For H excitation, $\sigma$(DCP) remained relatively flat, whereas for D excitation it exhibits periodic minima at intervals corresponding to (n+$\frac{1}{2}$)$\pi$ rotations from $\ket{\Uparrow}$. At these minima, the D-polarized pulse prepares the spin in the superpositions $\left|\Uparrow\Downarrow\uparrow\right\rangle$±$i\left|\Uparrow\Downarrow\downarrow\right\rangle$, whose projection onto the circular eigenbases is stationary under the applied magnetic field, and therefore the DCP contrast diminishes. The complementary behaviour of the two linear excitations provides high-contrast read-out of at least one spin quadrature at any moment in the precession cycle, a prerequisite for complete state reconstruction.

\begin{figure}[H]
\centering
\includegraphics[width=0.5\textwidth]{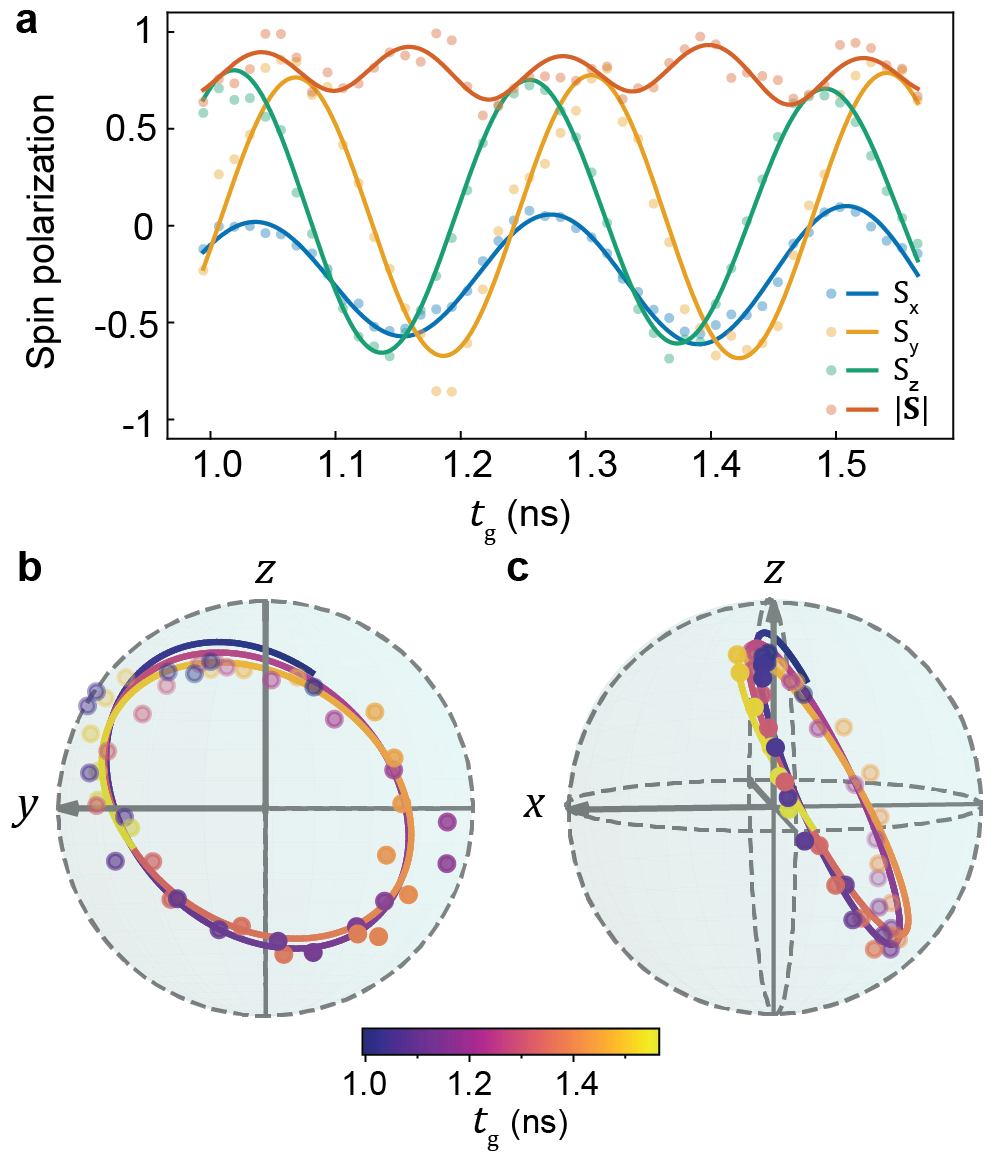}
\caption{\textbf{Tomographic reconstruction of ground state hole spin.} (a) Time evolution of the ground-state spin polarization components $S_x$ (blue), $S_y$ (yellow), $S_z$ (green), and the total spin magnitude $|\boldsymbol{S}|$ (red), as a function of the residence time in the ground state $t_{g}$, fitted with damped oscillator model. (b, c) Bloch sphere visualization of the reconstructed spin trajectories during coherent precession. (b) shows projection of the spin evolution onto the $y$–$z$ plane, while (c) reveals an inclination of the precession plane relative to the $y$–$z$ plane. Color gradient encodes elapsed $t_{g}$.}
\label{fig:tomo} 
\end{figure} 

Having established read-outs with complementary H/D-polarized excitation, we combine data from varying windows of ground-state precession intervals (e.g. $t_{g}$ = 1 to 1.55 ns) to reconstruct the evolving spin polarization \(\boldsymbol{S} = (S_x, S_y, S_z)\) of the hole. \cite{PhysRevB.105.L041407} For each spin precession time, we solve the coupled relations from simultaneous fits to the DCP time traces:
\begin{equation}
\left\{
\begin{aligned}
\quad A_0^{\mathrm{H}} &= \sqrt{S_y^{2} + S_z^{2}}, &\qquad
\varphi_0^{\mathrm{H}} &= \arctan\!\left(\dfrac{S_y}{S_z}\right) \\[4pt]
\quad A_0^{\mathrm{D}} &= \sqrt{S_x^{2} + S_z^{2}}, &\qquad
\varphi_0^{\mathrm{D}} &= \arctan\!\left(\dfrac{S_x}{S_z}\right).
\end{aligned}
\right.
\end{equation}

The reconstructed spin projections $S_x$, $S_y$, and $S_z$ are plotted in \autoref{fig:tomo}a. The common oscillation period of all three traces yields a hole $g$-factor of $g_{h}$=0.254±0.001, which is in excellent agreement with the value obtained in the previous section. The Bloch vector is well preserved during our probing window, giving an average spin purity $\langle$$|\boldsymbol{S}|$$\rangle$=0.79±0.11. The weak decay of the precession envelopes over the 600-ps acquisition window implies a lower bound of $T_{2}^{*}(h)>10$  ns for the hole's spin coherence time. A full evaluation of $T_{2}^{*}(h)$ would require assessing longer precession intervals, achievable by tuning delay times between the initialisation and probe PA pulses, thereby varying $t_{g}$ by more than the excited lifetime. The residual oscillation in $|\boldsymbol S|$ likely stems from a small mismatch between the nominal H/D polarization in the reference frame of the QD and the probe pulse polarisation, due to a fixed phase offset from $\pi/2$ or residual ellipticity; this causes the mapping to depart from the ideal $\pi/2$ rotation and imprints a phase dependent modulation. The spin trajectory across the Bloch sphere is provided in \autoref{fig:tomo}b,c. The $y$–$z$ projection (\autoref{fig:tomo}b) traces an elongated ellipse rather than a circle, while (\autoref{fig:tomo}c) reveals a $\sim18^\circ$ inclination of the precession plane relative to the idealized $y-z$ plane. Such off-axis precession can arise from an anisotropic hole $g$-tensor \cite{van2016anisotropy} or from a quasi-static Overhauser field generated by the quasi-static nuclear spin bath \cite{stockill2016quantum}. Other systematic offsets can also be introduced by a misalignment of the external Voigt field, or by mismatches between the preparation and analysis polarizations. For applications requiring in-plane precession (e.g., multiphoton entanglement), being able to identify such tilt through spin-tomography is essential as it introduces phase errors in the spin-photon interface.
Nevertheless, this misalignment could be counteracted by precision field steering, active polarization stabilization, and by optical nuclear spin cooling protocols that narrow the Overhauser distribution  \cite{gangloff2019quantum, nguyen2023enhanced}.

\section*{Conclusion}

In summary, we demonstrate an all-optical protocol that achieves complete quantum-state tomography of a heavy-hole spin in an InAs/InP QD emitting directly in the telecom C-band. PA excitation laser pulses with orthogonal linear polarizations provide complementary projection of the ground state phase onto the trion qubit, and time-tagged polarization correlation histograms enable reconstruction of the full Bloch vector, directly revealing the heavy-hole spin dynamics. From a single dataset we simultaneously verify the ratios of the hole and electron $g$-factors, extract lower bounds on their coherence. Using this tool we identify the spin with longer coherence and larger $g$-factor ratio as the more suitable photon-entangler qubit for deterministic multiphoton entanglement, without resorting to microwaves, electrical contacts, or polarization filtering.  In addition, the tomography analysis reveals spin purity evolution and provides crucial insight into the alignment between the precession plane and the spin axes, identifying anisotropy and phase offsets that would impact the fidelity of continuous Lindner-Rudolph cluster state generation. Crucially, the same tomographic reconstruction can be used constructively to infer the effective spin–photon mapping implemented and to compute an optimized excitation and measurement polarization basis that absorbs the phase offsets and improving the operational cluster-state fidelity under realistic misalignment, while the true decoherence still sets the ultimate limit. Taken together, our results provide informative diagnostics for optimising solid-state spin-photon interfaces en route to scalable cluster-state generation and fibre-based all-optical quantum repeater nodes.

\section*{Methods}
\textbf{Sample description:} 

Self-assembled InAs/InP quantum dots were grown by droplet epitaxy within a $3\lambda/2$ InP cavity formed by a 20-pair $\left((\mathrm{Al}_{0.30}\mathrm{Ga}_{0.70})_{0.48}\mathrm{In}_{0.52}\mathrm{As}\right)$/InP bottom DBR and a three-pair top DBR to enhance photon collection into the microscope. A 1~mm diameter cubic zirconia solid immersion lens was mounted on the sample to further improve photon collection efficiency.

\noindent \textbf{Experimental setup: } 

Phonon-assisted excitation is realised using a broadband pulsed C-band laser, with initial spectral filtering followed by temporal stretching to 15.8~ps using a 4-f spectral filter to produce wavelength-tunable pulses (FWHM 0.22~nm) optimised for efficient longitudinal acoustic PA excitation. The stretched pulses are routed through a fibre-based pulse sequence generator, employing non-polarising beamsplitters and precisely spliced fibre delays to produce two excitation pulses with precise time delays options within the 80~MHz repetition cycle (see Ref\cite{laccotripes2025entangled} for a detailed diagram of the setup).

The QD sample is cooled to 4~K in a closed-cycle cryostat, with excitation and photon collection performed using a free-space confocal microscope. The polarisation of the excitation pulses is controlled using a quarter and half wave plate pair, with a polarimeter used to align the laser polarisation to the QD reference frame. To isolate the QD emission, a cascade of three high-extinction bandpass filters is used to reject residual laser light.

For polarisation-resolving detection, electronic polarisation controllers and polarising beamsplitters are employed to project onto the desired polarisation bases, with photons detected using superconducting nanowire single-photon detectors with a system timing jitter of $\sim$ 40~ps.

\section*{Data Availability }
The authors declare that the data supporting the findings of this study are available from the corresponding authors upon reasonable request.

\bibliographystyle{naturemag} % Use for unsorted references
\bibliography{tomography}

\section*{Acknowledgements}
The authors gratefully acknowledge the usage of wafer material developed during earlier projects in partnership with the National Epitaxy Facility at the University of Sheffield. They further acknowledge funding from the Ministry of Internal Affairs and Communications, Japan, via the project `Research and Development for Construction of a Global Quantum Cryptography Network' in `R\&D of ICT Priority Technology' (JPMI00316).  P. L. gratefully acknowledges funding from the Engineering and Physical Sciences Research Council (EPSRC) via the Centre for Doctoral Training in Connected Electronic and Photonic Systems, grant EP/S022139/1. \\

\section*{Author Contributions Statement}
J.H and G.S devised the experiment. D.A.R, A.J.S, T.M, and R.M.S supervised the project. T.M and R.M.S guided the experiment. G.S fabricated the emission enhanced QD structure. J.H, G.S, and P.L performed the measurements and analysed the data. J.H, G.S, P.L and T.M wrote the manuscript, with all authors contributed to discussing the results and commenting on the manuscript.\\

\section*{Competing Interests Statement}
The authors declare that they have no competing financial interests.

\end{document}